\documentstyle[preprint,prl,aps]{revtex}

\newcommand{\be}{\begin{equation}}
\newcommand{\ee}{\end{equation}}
\newcommand{\ba}{\begin{eqnarray}}
\newcommand{\ea}{\end{eqnarray}}

\draft

\begin{document}
\title{Tsallis' entropy maximization procedure revisited}
\author{S. Mart\'{\i}nez${^{1,\,2}}$\thanks{%
E-mail: martinez@venus.fisica.unlp.edu.ar}, F. Nicol\'{a}s, F. Pennini${%
^{1,\,2}}$\thanks{
E-mail: pennini@venus.fisica.unlp.edu.ar}, and A. Plastino${^{1,\,2}}$%
\thanks{
E-mail: plastino@venus.fisica.unlp.edu.ar}}
\address{$1$ Physics Department, National University La Plata, C.C. 727 ,\\
1900 La Plata, Argentina \\ $^2$ Argentine National Research
Council (CONICET)} \maketitle

\begin{abstract}
The proper way of averaging is an important question with regards to
Tsallis' Thermostatistics. Three different procedures have been thus far
employed in the pertinent literature. The third one, i.e., the
Tsallis-Mendes-Plastino (TMP) \cite{mendes} normalization procedure,
exhibits clear advantages with respect to earlier ones. In this work, we
advance a distinct (from the TMP-one) way of handling the Lagrange
multipliers involved in the extremization process that leads to Tsallis'
statistical operator. It is seen that the new approach considerably
simplifies the pertinent analysis without losing the beautiful properties of
the Tsallis-Mendes-Plastino formalism.


\end{abstract}

\newpage

\section{Introduction}

Tsallis' thermostatistics \cite{mendes,t01,t1,t03,t04,t05} is by now
recognized as a new paradigm for statistical mechanical considerations. One
of its crucial ingredients, Tsallis' normalized probability distribution
\cite{mendes}, is obtained by following the well known MaxEnt route \cite
{katz}. One maximizes Tsallis' generalized entropy \cite{t1,t2}
\begin{equation}
\frac{S_{q}}{k}=\frac{1-\sum_{i=1}^{w}p_{i}^{q}}{q-1},  \label{entropia}
\end{equation}
($k\equiv k(q)$ tends to the Boltzmann constant $k_{B}$ in the limit $%
q\rightarrow 1$ \cite{t01}) subject to the constraints (generalized
expectation values) \cite{t01}

\begin{eqnarray}
\sum_{i=1}^wp_i &=&1 \\
\frac{\sum_{i=1}^wp_i^qO_j^{(i)}}{\sum_{i=1}^wp_i^q} &=&\left\langle
\left\langle O_j\right\rangle \right\rangle _q,  \label{vinculos}
\end{eqnarray}
where $p_i$ is the probability assigned to the microscopic configuration $i$
($i=1,\ldots ,w$) and one sums over all possible configurations $w$. $%
O_j^{(i)}$ ($j=1,\ldots ,n$) denote the $n$ relevant observables (the
observation level \cite{aleman}), whose generalized expectation values $%
\left\langle \left\langle O_j\right\rangle \right\rangle _q$ are (assumedly)
a priori known.

The Lagrange multipliers recipe entails maximizing \cite{mendes}
\begin{equation}
F=\frac{S_{q}}{k}-\alpha _{0}\left( \sum_{i=1}^{w}p_{i}-1\right)
-\sum_{j=1}^{n}\lambda _{j}\left( \frac{\sum_{i=1}^{w}p_{i}^{q}O_{j}^{(i)}}{%
\sum_{i=1}^{w}p_{i}^{q}}-\left\langle \left\langle O_{j}\right\rangle
\right\rangle _{q}\right) ,  \label{F}
\end{equation}
yielding

\begin{equation}
p_{i}=\frac{f_{i}^{1/(1-q)}}{\bar{Z}_{q}},  \label{pv}
\end{equation}
where

\begin{equation}
f_{i}=1-\frac{(1-q)\sum_{j}\lambda _{j}\left( O_{j}^{(i)}-\left\langle
\left\langle O_{j}\right\rangle \right\rangle _{q}\right) }{\sum_{j}p_{j}^{q}%
},  \label{fin}
\end{equation}
is the so-called configurational characteristic \cite{pennini1} and
\begin{equation}
\bar{Z}_{q}=\sum_{i}f_{i}^{1/(1-q)},  \label{Zq}
\end{equation}
stands for the partition function.

The above procedure, originally employed in \cite{mendes}, overcomes most of
the problems posed by the old, unnormalized way of evaluating
Tsallis'generalized mean values \cite{mendes,pennini}. Some hardships
remain, though. One of them is that numerical difficulties are sometimes
encountered, as the $p_{i}$ expression is {\it explicitly self-referential}.
An even more serious problem is also faced: a maximum is not necessarily
guaranteed. Indeed, analyzing the concomitant Hessian so as to ascertain
just what kind of extreme we face, one encounters the unpleasant fact that
this Hessian is not diagonal.

In the present effort we introduce an alternative Lagrange route, that
overcomes the above mentioned problems.

\section{The new Lagrange multipliers' set}

We extremize again (\ref{entropia}) subject to the constraints (\ref
{vinculos}), but, and herein lies the central idea, {\it rephrase} (\ref{F})
by recourse to the alternative form

\begin{equation}
\sum_{i=1}^w p_i^q\left( O_j^{(i)} - \left<\left<O_j\right>\right>_q\right)
= 0,  \label{vinculosnuevos}
\end{equation}
with $j=1,\ldots,n$. We have now

\begin{equation}
F=\frac{S_{q}}{k}-\alpha _{0}\left( \sum_{i=1}^{w}p_{i}-1\right)
-\sum_{j=1}^{n}\lambda _{j}^{\prime }\sum_{i=1}^{w}p_{i}^{q}\left(
O_{j}^{(i)}-\left\langle \left\langle O_{j}\right\rangle \right\rangle
_{q}\right) ,  \label{Fnueva}
\end{equation}
so that, following the customary variational procedure and eliminating $%
\alpha _{0}$ we find that the probabilities are, formally, still given by (%
\ref{pv}). However, in terms of the new set of Lagrange multipliers, the
configurational characteristics {\it do not depend explicitly on the
probabilities}

\begin{equation}
f_{i}^{\prime }=1-(1-q)\sum_{j}\lambda _{j}^{\prime }\left(
O_{j}^{(i)}-\left\langle \left\langle O_{j}\right\rangle \right\rangle
_{q}\right) .  \label{finueva}
\end{equation}

Comparing (\ref{finueva}) with (\ref{fin}), it is clear that the Lagrange
multipliers $\lambda _{j}$ of the Tsallis-Mendes-Plastino formalism (TMP)
\cite{mendes} and $\lambda _{j}^{\prime }$ (present treatment) can be
connected via

\begin{equation}
\lambda _j^{\prime }=\frac{\lambda _j}{\sum_ip_i^q},  \label{conectl}
\end{equation}
which leads to the nice result $f_i^{\prime }=f_i$. The probabilities that
appear in (\ref{conectl}) are {\it those special ones that maximize the
entropy}, not generic ones. The ensuing, new partition function is also of
the form (\ref{Zq}), with $f_i>0$ the well known Tsallis' cut-off condition
\cite{t1,t2}. Notice that now the expression for the MaxEnt probabilities $%
p_i$ is NOT explicitly self-referential.

In order to ascertain the kind of extreme we are here facing we study the
Hessian, that now is of {\it diagonal} form. The maximum condition
simplifies then to the requirement

\begin{equation}
\frac{\partial ^2F}{\partial p_i^2}<0.
\end{equation}

The above derivatives are trivially performed yielding

\begin{equation}
\frac{\partial ^{2}F}{\partial p_{i}^{2}}=-qp_{i}^{q-2}f_{i},
\end{equation}
which formally coincides with the maximum requirement one finds in the case
of Tsallis' unnormalized formalism. Since the $f_{i}$ are positive-definite
quantities, for a maximum one should demand that $q>0$.

Extremes found by following the celebrated Lagrange procedure depend only on
the nature of the constraints, not on the form in which they are expressed.
Thus, the two sets of multipliers lead to the same numerical values for the
micro-state probabilities. Via (\ref{conectl}) one is always able to
establish a connection between both treatments.

The present algorithm exhibits the same nice properties of the TMP
formalism, namely:

\begin{itemize}
\item  The MaxEnt probabilities are invariant under uniform shifts of the
Hamiltonian's energy spectrum (see, for instance, the illuminating
discussion of Di Sisto et al.\cite{disisto}).

Indeed, after performing the transformation

\begin{eqnarray}
\epsilon _{i} &\rightarrow &\epsilon _{i}+\epsilon _{0} \\
U_{q} &\rightarrow &U_{q}-\epsilon _{0},
\end{eqnarray}
on equation (\ref{pv}), with $f_{i}$ given by (\ref{finueva}), we trivially
find that the probabilities $p_{i}$ keep their forms invariant if the $%
\lambda _{j}^{\prime }$ do not change. Due to relation (\ref{conectl}), the $%
\lambda _{j}$ are invariant too.

\item  The mean value of unity equals unity, i.e., $\left\langle
\left\langle 1\right\rangle \right\rangle _{q}=1$, which is not the case
with the unnormalized expectation values \cite{t1,t2}.

\item  One easily finds that, for two independent subsystems $A,\,\,\,B$,
energies add up: $U_{q}(A+B)=U_{q}(A)+U_{q}(B)$.
\end{itemize}

\section{OLM treatment and R\`{e}nyi's measure}

The above OLM recipe has the purpose of diagonalizing
the Hessian characterizing the kind of extreme one
finds after a Lagrange treatment. It is natural to ask
whether the concomitant recipe works also for other
information measures in addition to the Tsallis' one.
We discuss next the R\`{e}nyi's measure ($R_{q}$)
 instance \cite{pennini,renyi,beck,lenzi} in an OLM context.
 We have \cite{t06}
\begin{equation}
\frac{R_{q}}{k}=\frac{\ln \left[ 1+(1-q)S_{q}\right] }{1-q},
\end{equation}
or (Cf. (\ref{entropia}))

\begin{equation}
\frac{R_{q}}{k}=\frac{{\rm ln}\left( \sum_{i}p_{i}^{q}\right) }{q-1}.
\label{SR}
\end{equation}

We extremize now $%
R_{q}$ subject to the constraints (\ref {vinculosnuevos}) with
Lagrange multipiers $\mu_j^*$. The ensuing Lagrange functional
reads

\begin{equation}
F^{*}=\frac{R_{q}}{k}-\alpha _{0}\left(
\sum_{i=1}^{w}p_{i}-1\right) -\sum_{j=1}^{n}\mu
_{j}^{*}\sum_{i=1}^{w}p_{i}^{q}\left( O_{j}^{(i)}-\left\langle
\left\langle O_{j}\right\rangle \right\rangle _{q}\right) ,
\end{equation}
whose first derivatives are
\begin{equation}
\frac{\partial F^{*}}{\partial p_{i}}=\frac{qp_{i}^{q-1}}{%
(q-1)\sum_{i}p_{i}^{q}}-\alpha _{0}-qp_{i}^{q-1}\sum_{j}\mu
_{j}^{*}\left( O_{j}^{(i)}-\left\langle \left\langle
O_{j}\right\rangle \right\rangle _{q}\right) =0.  \label{Fp}
\end{equation}

The entropy term will lead to a self referencial probability
distribution.  We have

\begin{equation}
p_{i}=\frac{(f_{i}^{*})^{\frac{1}{1-q}}}{Z_{q}^{*}},
\end{equation}
with a functional characteristic given by
\begin{equation}
f_{i}^{*}=1+(1-q)\sum_{i}p_{i}^{q}\sum_{j}\mu _{j}^{*}\left(
O_{j}^{(i)}-\left\langle \left\langle O_{j}\right\rangle
\right\rangle _{q}\right) , \label{f*}
\end{equation}
and the partition function
\begin{equation}
Z_{q}^{*}=\sum_{i}(f_{i}^{*})^{\frac{1}{1-q}}.
\end{equation}

We see that the OLM treatment leads to a self-referential solution
that the ``R\`enyi-TMP''  treatament does not exhibit. Indeed, on
account of the fact both the constraints and the entropy share the
$\sum_{i}p_{i}^{q}$ denominator, the TMP-R\`enyi distribution has
characteristics of the form \cite{lenzi}
\begin{equation}
f_{i}=1+(1-q)\sum_{j}\mu _{j}\left( O_{j}^{(i)}-\left\langle
\left\langle O_{j}\right\rangle \right\rangle
_{q}\right).\label{ff}
\end{equation}

Comparing Eqs. (\ref{f*}) and (\ref{ff}), one observes that the
Lagrange multipliers are connected via

\be
\mu _{j}^{*}=\frac{\mu _{j}}{\sum_{i}p_{i}^{q}}, \ee that is the
same relation found for Tsallis' measure (see Equation
(\ref{conectl})). The following correspondence can be appreciated
$$ \lambda_j' = \mu_j,$$ since the configurational characteristics
of Equation (\ref{ff}) is identical to the one of Equation
(\ref{finueva}).

Consider now, for  the second derivatives, the  non-diagonal
terms. One has. in the OLM instance,
\begin{equation} \label{ecu1}
\frac{\partial ^{2}F^{*}}{\partial p_{j}\partial p_{i}}=\frac{q^{2}}{1-q}%
\frac{(p_{i}p_{j})^{q-1}}{\left( \sum_{k}p_{k}^{q}\right) ^{2}}.
\end{equation}

If one uses  the standard TMP approach of Ref. \cite{mendes} one
finds, instead, for these non-diagonal terms,
\begin{equation} \label{ecu2}
\frac{\partial ^{2}F}{\partial p_{j}\partial p_{i}}=\frac{q^{2}}{1-q}\frac{%
f_{i}(p_{i}p_{j})^{q-1}}{\left( \sum_{k}p_{k}^{q}\right) ^{2}}.
\end{equation}

There is an important distinction to be made between
(\ref{ecu1}) and (\ref{ecu2}), that refers to the {\it
origin} of the non-diagonal terms. In the first case,
they have as a source just the entropy term of the
Lagrangian. In the second, the constraints also
contribute.

As a consequence we gather that the OLM approach will work as
nicely as in the Tsallis case only for those measures that do not
give rise to off-diagonal terms in the associated Hessian.

\section{Thermodynamics}

We pass now to the question of writing down the basic mathematical
relationships of Thermodynamics, as expressed with respect to the new set of
Lagrange multipliers $\lambda _j^{\prime }$.

In order to do this in the most general quantal fashion we shall work in a
basis-independent way. This requires consideration of the statistical
operator (or density operator) $\hat{\rho}$ that maximizes Tsallis' entropy,
subject to the foreknowledge of $M$ generalized expectation values
(corresponding to $M$ operators $\widehat{O}_j$). These take the form
\begin{equation}
\left\langle \left\langle \widehat{O}_j\right\rangle \right\rangle _q=\frac{%
Tr(\hat{\rho}^q\widehat{O}_j)}{Tr(\hat{\rho}^q)},\qquad j=1,...,M.
\label{vm}
\end{equation}

To these we must add, of course, the normalization requirement
\begin{equation}
Tr\hat{\rho}=1.  \label{norm}
\end{equation}

The TMP formalism, where relations are written in terms of the ``old''
Lagrange multipliers $\lambda _j$, yields the usual thermodynamical
relationships \cite{mendes}, namely
\begin{eqnarray}
\frac{\partial }{\partial \left\langle \left\langle \widehat{O}%
_j\right\rangle \right\rangle _q}\left(\frac{S_q}{k} \right) &=&\lambda _j  \label{term1} \\
\frac \partial {\partial \lambda _i}\left( \ln _qZ_q\right) &=&-\left\langle
\left\langle \widehat{O}_j\right\rangle \right\rangle _q,  \label{term2}
\end{eqnarray}
where
\begin{equation}
{\rm \ln }_q\bar{Z}_q=\frac{\bar{Z}_q^{1-q}-1}{1-q}  \label{lnq}
\end{equation}
and \cite{mendes}
\begin{equation}
\ln _qZ_q={\rm \ln }_q\bar{Z}_q-\sum_j\lambda _j\ \left\langle \left\langle
\widehat{O}_j\right\rangle \right\rangle _q,  \label{lnqzq}
\end{equation}
so that the essential mathematical structure of Thermodynamics is preserved.

Following the standard procedure \cite{t04,t06} one gets
\begin{equation}
\hat{\rho}=\bar{Z}_q^{-1}\left[ 1-(1-q)\sum_j^M\,\lambda _j^{\prime }\left(
\widehat{O}_j-\left\langle \left\langle \widehat{O}_j\right\rangle
\right\rangle _q\right) \right] ^{\frac 1{1-q}},  \label{rho}
\end{equation}
where $\bar{Z}_q$ stands for the partition function

\begin{equation}
\bar{Z}_q=Tr\left[ 1-(1-q)\sum_j\lambda _j^{\prime }\left( \widehat{O}%
_j-\left\langle \left\langle \widehat{O}_j\right\rangle \right\rangle
_q\right) \right] ^{\frac 1{1-q}}.  \label{Zqp}
\end{equation}

Enters here Tsallis'cut-off condition \cite{t04,t06}. The form (\ref{rho})
does not a priori guarantee that we will have a positive-definite operator.
Some additional considerations are requested.

Consider the operator

\begin{equation}
\widehat{A}\ =1-(1-q)\sum_j\lambda _j^{\prime }\left( \widehat{O}%
_j-\left\langle \left\langle \widehat{O}_j\right\rangle \right\rangle
_q\right)  \label{A}
\end{equation}
enclosed within parentheses in (\ref{rho}). One must ensure its
positive-definite character. This entails that the eigenvalues of $\hat{A}$
must be non-negative quantities. This can be achieved by recourse to an
heuristic cut-off procedure. We replace (\ref{rho}) by
\begin{equation}
\hat{\rho}=\bar{Z}_q^{-1}\left[ \hat{A}\ \Theta (\hat{A})\right] ^{1/(1-q)},
\label{A1}
\end{equation}
with $\bar{Z}_q$ given by
\begin{equation}
\bar{Z}_q=Tr\left[ \hat{A}\ \Theta (\hat{A})\right] ^{1/(1-q)},  \label{A2}
\end{equation}
where $\Theta (x)$ is the Heaviside step-function. Equations (\ref{A1})-(\ref
{A2}) are to be re-interpreted as follows. Let $\left| i\right\rangle $ and $%
\alpha _i,$ stand, respectively, for the eigenvectors and eigenvalues of the
operator (\ref{A}), whose spectral decomposition is then
\begin{equation}
\hat{A}=\sum_i\alpha _i\ \left| i\right\rangle \left\langle i\right| .
\end{equation}

In the special basis used above $\hat{\rho}\ $ adopts the appearance
\begin{equation}
\hat{\rho}=\bar{Z}_q^{-1}\sum_if(\alpha _i)\ \left| i\right\rangle
\left\langle i\right| ,
\end{equation}
with $f(x)$ defined as
\begin{equation}
f(x)=0,\,\,\,for\,\,\,x\leq 0,
\end{equation}
and
\begin{equation}
f(x)=x^{\frac 1{1-q}},\,\,\,for\,\,\,x>0.
\end{equation}

Notice that $f(x)$ possesses, for $0<q<1,$ a continuous derivative for all $%
x.$ Moreover,
\begin{equation}
\frac{df(x)}{dx}=\left( \frac{1}{q-1}\right) \left[ x\Theta (x)\right] ^{%
\frac{q}{1-q}}.  \label{df}
\end{equation}

In terms of the statistical operator, Tsallis' entropy $S_{q}$ reads
\begin{eqnarray}
\frac{S_{q}}{k} &=&\frac{1}{q-1}\ Tr\left[ \hat{\rho}^{q}\left( \hat{\rho}%
^{1-q}-\hat{I}\right) \right]  \nonumber \\
&=&\frac{1}{q-1}\ Tr\left[ \hat{\rho}^{q}\left( \bar{Z}_{q}^{q-1}\hat{A}\
\Theta (\hat{A})-\hat{I}\right) \right] \\
&=&\frac{\bar{Z}_{q}^{q-1}}{q-1}\ Tr\left[ \hat{\rho}^{q}\hat{A}\ \Theta (%
\hat{A})\right] -\frac{Tr(\hat{\rho}^{q})}{(q-1)},  \nonumber
\end{eqnarray}
where $\hat{I}$ is the unity operator.

Obviously, $\hat{\rho}$ commutes with $\hat{A}$. The product of these two
operators can be expressed in the common basis that diagonalizes them

\begin{equation}
\hat{\rho}^q\,\hat{A}\,\Theta (\hat{A})=\bar{Z}_q^{-q}\,\sum_i\,[f(\alpha
_i)]^q\,\alpha _i|i\rangle \langle i|,
\end{equation}
which entails, passing from the special basis $|i\rangle $ to the general
situation, that
\begin{equation}
\hat{\rho}^q\,\hat{A}\,\Theta (\hat{A})=\hat{\rho}^q\left[
1-(1-q)\sum_j\lambda _j^{\prime }\left( \widehat{O}_j-\left\langle
\left\langle \widehat{O}_j\right\rangle \right\rangle _q\right) \right] ,
\end{equation}
and, consequently
\begin{equation}
\frac{S_q}{k}=\frac{\bar{Z}_q^{q-1}-1}{q-1}\ Tr\left( \hat{\rho}^q\right) +\bar{Z}%
_q^{q-1}\sum_j\lambda _j^{\prime }\ Tr\left[ \hat{\rho}^q\ \left( \widehat{O}%
_j-\left\langle \left\langle \widehat{O}_j\right\rangle \right\rangle
_q\right) \right] .
\end{equation}

Since the last term of the right-hand-side vanishes, by definition (\ref
{vinculosnuevos}), we finally arrive at
\begin{equation}
\frac{S_{q}}{k}=\frac{\bar{Z}_{q}^{q-1}-1}{q-1}\ Tr\left( \hat{\rho}%
^{q}\right) .  \label{Sq}
\end{equation}

Now, from the very definition (in terms of $\hat \rho $) of Tsallis' entropy
$S_q$ \cite{t04,t06}, we find
\begin{equation}
Tr\left( \hat{\rho}^q\right) =1+(1-q)\frac{S_q}k ,  \label{Sqq}
\end{equation}
so that (\ref{Sq}) and (\ref{Sqq}) lead to
\begin{equation}
Tr(\hat{\rho}^q)=\bar{Z}_q^{1-q}  \label{relac1}
\end{equation}
and
\begin{equation}
S_q=k\;{\rm \ln }_q\bar{Z}_q,  \label{S2}
\end{equation}
where ${\rm \ln }_q\bar{Z}_q$ has been introduced in (\ref{lnq}).

Using (\ref{relac1}), equation (\ref{conectl}) can be rewritten as

\begin{equation}  \label{lambda'}
\lambda_j^{\prime }= \frac{\lambda_j}{\bar{Z}_{q}^{1-q}}.
\end{equation}

Following \cite{mendes} we define now
\begin{equation}
\ln _qZ_q^{\prime }={\rm \ln }_q\bar{Z}_q-\sum_j\lambda _j^{\prime }\
\left\langle \left\langle \widehat{O}_j\right\rangle \right\rangle _q,
\label{lnqz'}
\end{equation}
which leads finally to (see (\ref{term1}) and (\ref{term2}))
\begin{eqnarray}
\frac{\partial}{\partial \left\langle \left\langle \widehat{O}%
_j\right\rangle \right\rangle _q} \left(\frac{S_q}{k}\right)&=&
\bar{Z}_q^{1-q}\lambda _j^{\prime}=\lambda_j
\label{termo1} \\
\frac \partial {\partial \lambda _j^{\prime }}\left( \ln _qZ_q^{\prime
}\right) &=&-\left\langle \left\langle \widehat{O}_j\right\rangle
\right\rangle _q.  \label{termo2}
\end{eqnarray}

Equations (\ref{termo1}) and (\ref{termo2}) constitute the basic
Information Theory relations on which to build up, {\em \`a la}
Jaynes \cite{katz}, Statistical Mechanics. Notar que OLM conduce
a las mismas relaciones que TMP, y que no ha sido recesario tomar
$k$ como constante en ninguna parte del desarrollo.

As a special instance of Eqs. (\ref{termo1}) and (\ref{termo2}) let us
discuss the Canonical Ensemble, where they adopt the appearance
\begin{eqnarray}
\frac{\partial }{\partial
U_q}\left(\frac{S_q}{k}\right)&=&\bar{Z}_q^{1-q}\beta ^{\prime }
=\beta \label{C1} \\
\frac \partial {\partial \beta ^{\prime }}\left( \ln _qZ_q^{\prime }\right)
&=&-U_q,  \label{C2}
\end{eqnarray}
where (see equation (\ref{lnqz'}))
\begin{equation}
\ln _qZ_q^{\prime }=\ln _q\bar{Z}_q-\beta ^{\prime }U_q.  \label{lnqz'can}
\end{equation}

Proponemos que $\beta^{\prime}$ Finally, the specific heat reads

\begin{equation}
C_q=\frac{\partial U_q}{\partial T}=\frac{\partial
U_q}{\partial
\beta'}\frac{d\beta'}{dT}=-\frac{\partial U_q}{\partial
\beta'}\frac{1}{k'T^2}=
-k^{\prime }\beta ^{\prime 2}\frac{%
\partial U_q}{\partial \beta ^{\prime }}.
\end{equation}

We conclude that the mathematical form of the thermodynamic relations is
indeed preserved by the present treatment. Both sets of Lagrange multipliers
accomplish this feat and they are connected via (\ref{conectl}). The primed
one, however, allows for a simpler treatment, as will be illustrated below.

\section{Simple applications}

We consider now some illustrative examples. They are chosen in such a manner
that each of them discusses a different type of situation: classical and
quantal systems, the latter in the case of both finite and infinite number
of levels.

\subsection{The classical harmonic oscillator}

\label{cho}

Let us consider the classical harmonic oscillator in the canonical ensemble.
We can associate with the classical oscillator a continuous energy spectrum $%
\epsilon(n)= \epsilon \; n $ with $\epsilon>0$ and $n \in {\cal R}^+$
compatible with the cut-off condition. The ensuing MaxEnt probabilities
adopt the appearance

\begin{equation}
p_{q}(n,t^{\prime })=\frac{\left[ f_{q}(n,t^{\prime })\right] ^{1/(1-q)}}{%
\bar{Z}_{q}(t^{\prime })},  \label{pq}
\end{equation}
where

\begin{equation}
f_{q}(n,t^{\prime })=1-(1-q)\frac{(n-u_{q})}{t^{\prime }},  \label{fq}
\end{equation}
and

\begin{equation}
\bar{Z}_q(t^{\prime })=\int_0^{n_{max}}\left[ f_q(n,t^{\prime })\right]
^{1/(1-q)}dn,  \label{zq}
\end{equation}
with $u_q=U_q/\epsilon $ and $t^{\prime }=k^{\prime }T/\epsilon $. We have
introduced also $n_{max}$ as the upper integration limit on account of
Tsallis' cut-off condition. One appreciates the fact that $%
n_{max}\rightarrow \infty $ if $q>1$. $n_{max}$ is, of course, the maximum $%
n $-value that keeps $\left[ f_q(n,t^{\prime })\right] ^{1/(1-q)}>0$ for $%
q<1 $.

The normalization condition reads

\begin{equation}
u_{q}(t^{\prime })=\frac{\int_{0}^{n_{max}}\left[ p_{q}(n,t^{\prime
})\right] ^{q}n\;dn}{\int_{0}^{n_{max}}\left[ p_{q}(n,t^{\prime })\right]
^{q}dn},
\end{equation}
or, using (\ref{pq}),

\begin{equation}
u_{q}(t^{\prime })=\frac{\int_{0}^{n_{max}}\left[ f_{q}(n,t^{\prime
})\right] ^{q/(1-q)}n\;dn}{\int_{0}^{n_{max}}\left[ f_{q}(n,t^{\prime
})\right] ^{q/(1-q)}dn}.  \label{uq}
\end{equation}

Due to the form of $f_{q}$, equation (\ref{uq}) constitutes a well-defined
expression. By explicitly performing the integrals for $1<q<2$ (for $q\geq 2$
the integrals diverge) we obtain

\begin{equation}
u_{q}(t^{\prime })=\frac{t^{\prime 2}/(2-q)(1+(1-q)u_{q}/t^{\prime
})^{(2-q)/(1-q)}}{t^{\prime }(1+(1-q)u_{q}/t^{\prime })^{1/(1-q)}}.
\end{equation}

After a little algebra, the above equation leads to the simple result

\begin{equation}
u_{q}(t^{\prime })=t^{\prime }.  \label{uc}
\end{equation}

Replacing now $u_q=U_q/\epsilon $ and $t^{\prime }=k^{\prime }T/\epsilon $,
we obtain $U_q=k^{\prime }T$, so that the specific heat reads

\begin{equation}
C_{q}=k^{\prime }.  \label{ceoc}
\end{equation}

It is worthwhile to remark that, in the case of this particular example, we
{\it formally} regain the usual expressions typical of the $q=1$ case. Due
to fact that we possess a degree of freedom in the definition of $k^{\prime}$%
, we can set $k^{\prime}=k_B$ and thus recover Gibbs' Thermodynamics.
Performing the pertinent integral and using (\ref{uc}), the partition
function becomes

\begin{equation}
\bar{Z}_q(t^{\prime})= t^{\prime} (2-q)^{1/(1-q)}.
\end{equation}

According to equation (\ref{lambda'}), $t^{\prime}$ can be written in terms
of $t$ and $\bar{Z}_q$, allowing us to recover \cite{mendes}
\begin{equation}
\bar{Z}_q(t)= t^{1/q} (2-q)^{1/[q(1-q)]},
\end{equation}
and, consequently,
\begin{eqnarray}
u_q&=& t^{1/q}(2-q)^{1/q} \\
C_q&=& \frac{k}{2} (2-q)^{1/q} t^{(1-q)/q}.
\end{eqnarray}

These results are identical to those of \cite{mendes}, but are here derived
in a remarkably {\it simpler} fashion.

\subsection{The two-level system and the quantum harmonic oscillator}

Let us consider the discrete case of a single particle with an energy
spectrum given by $E_n=\epsilon n$, where $\epsilon>0$ and $n=0,1,...,N$. If
$N=1$, we are facing the non degenerate two level system, while, if $n
\rightarrow \infty$, the attendant problem is that of the quantum harmonic
oscillator.

The micro-state probabilities are of the form, once again

\begin{equation}
p_n = \frac{f_n^{1/(1-q)}}{\bar{Z}_q}  \label{pi}
\end{equation}
with

\begin{equation}
\bar{Z}_q=\sum_{n=0}^Nf_n{}^{1/(1-q)}.  \label{zz}
\end{equation}

The configurational characteristics take the form

\begin{equation}
f_n(t^{\prime})=1-(1-q)(n-u_q)/t^{\prime}
\end{equation}
where again (see (\ref{cho})), $t^{\prime}=k^{\prime} T/\epsilon$ and $%
u_q=U_q/k^{\prime}$.

Using (\ref{pi}), the mean energy can be written as

\begin{equation}
u_q=\frac{\sum_{n=0}^Nf_n^{q/(1-q)}n}{\sum_{n=0}^Nf_n^{q/(1-q)}},
\end{equation}
which, using the explicit form of $f_n$ and rearranging terms, allows one to
write down the following equation
\begin{equation}
\sum_{n=0}^N\left[ 1-\frac{(1-q)}{t^{\prime }}(n-u_q)\right]
^{q/(1-q)}(n-u_q)=0,  \label{unique}
\end{equation}
which implicitly defines $u_q$. Notice that one does not arrive to a closed
expression. However, in order to numerically solve for $u_q$, we just face (%
\ref{unique}). This equation is easily solved by recourse to the so-called
``seed'' methods (cut-off always taken care of), with quick convergence
(seconds). This is to be compared to the TMP instance \cite{mendes}. In
their case, one faces a non-linear coupled system of equations in order to
accomplish the same task. This coupled system can be recovered from (\ref
{unique}) and (\ref{pi}), writing $t^{\prime }$ in terms of $t$.

\subsection{Magnetic Systems}

Consider now a very simple magnetic model, discussed, for instance, in \cite
{portesi}: a quantum system of $N$ spin 1/2 non-interacting atoms in the
presence of a uniform, external magnetic field $\vec{H}=H\hat{k}$ (oriented
along the unit vector $\hat{k}$). Each atom is endowed with a magnetic
moment $\widehat{\vec{\mu}}^{(i)}=g\,\mu _0\,\widehat{\vec{S}}^{(i)},$ where
$\mu _0=e/(2mc)$ is Bohr's magneton and $\widehat{\vec{S}}^{(i)}=(\hbar /2)\,%
\widehat{\vec{\sigma}}^{(i)}$, with $\widehat{\vec{\sigma}}^{(i)}$ standing
for the Pauli matrices. The concomitant interaction energy reads

\begin{equation}
\hat{{\cal H}}=-\sum_{i=1}^N\widehat{\vec{\mu}}^{(i)}\cdot \vec{H}=-\frac{%
g\mu _0}\hbar H\,\widehat{S}_z,
\end{equation}
where $\widehat{\vec{S}}=\sum_{i=1}^N\widehat{\vec{S}}^{(i)}$ the total
(collective) spin operator. The simultaneous eigenvectors of $\widehat{\vec{S%
}}^2$ and $\widehat{S}_z$ constitute a basis of the concomitant $2^N$%
-dimensional space. We have $\left| S,M\right\rangle $, with $S=\delta
,\ldots ,N/2,$ $M=-S,\ldots ,S,$ and $\delta \equiv N/2-[N/2]=0$ $(1/2)$ if $%
N$ is even (odd). The corresponding multiplicities are $%
Y(S,M)=Y(S)=N!(2S+1)/[(N/2-S)!(N/2+S+1)!]$ \cite{portesi}. We recast the
Hamiltonian in the simple form

\begin{equation}  \label{H}
\hat {{\cal H}}= - \frac{x^{\prime}}{\beta^{\prime}} \hat S_z,
\end{equation}
with $x^{\prime}=g\mu _{0}H\beta^{\prime} /\hbar $. Our statistical operator
can be written as

\begin{equation}
\hat{\rho}=\frac 1{\bar{Z}_q}\left[ 1-(1-q)x^{\prime }\left( \hat{S}%
_z-\left\langle \left\langle \hat{S}_z\right\rangle \right\rangle _q\right)
\right] ^{1/(1-q)},  \label{rho2}
\end{equation}
where

\begin{equation}  \label{Z2}
\bar{Z}_q = Tr \left[ 1-(1-q) x^{\prime} \left(\hat S_z- \left<\left< \hat S%
_z \right>\right>_q \right)\right]^{1/(1-q)}.
\end{equation}

Due to the cut-off condition, $1-(1-q) x^{\prime} \left(\hat S_z-
\left<\left< \hat S _z \right>\right>_q \right)>0$.

The mean value of the spin $z$-component is computed according to (\ref{vm})

\begin{equation}  \label{Sz2}
\left<\left< \hat S_z \right>\right>_q = \frac{Tr\left( \hat \rho^q \hat S_z
\right)}{Tr\left( \hat \rho^q \right)},
\end{equation}
so that, replacing (\ref{rho2}) into (\ref{Sz2}) and rearranging then terms
we arrive at

\begin{equation}
Tr \left\{ \left[ 1+(1-q) x^{\prime} \left(\hat S_z - \left<\left<\hat S%
_z\right>\right>_q\right)\right]^{q/(1-q)}\left(\hat S_z - \left<\left<\hat S%
_z\right>\right>_q\right)\right\} =0.
\end{equation}

More explicitly, one has

\begin{equation}
\sum_{S=\delta }^{N/2}Y(S)\sum_{M=-S}^{S}\left[ 1+(1-q)x^{\prime }\left(
M-\left\langle \left\langle \hat{S}_{z}\right\rangle \right\rangle
_{q}\right) \right] ^{q/(1-q)}\left( M-\left\langle \left\langle \hat{S}%
_{z}\right\rangle \right\rangle _{q}\right) =0,  \label{magne}
\end{equation}
which is the equation to be solved in order to find $\left\langle
\left\langle \hat{S}_{z}\right\rangle \right\rangle _{q}$.

Notice that, once again, one faces just a {\it single} equation, that can be
easily tackled. If one uses instead the TMP prescription (as discussed in
\cite{nosotros}) one has to solve a {\it coupled}, highly non-linear system
of equations. Such a system can be recovered from (\ref{magne}) if one
replaces $x^{\prime }$ by $x/Tr(\rho ^q)$ and adds the condition $Tr(\rho
^q) $ from (\ref{rho2}).

As in \cite{nosotros}, we consider now two asymptotic situations from the
present viewpoint.

For $x^{\prime }\rightarrow 0$ we Taylor-expand (\ref{magne}) around $%
x^{\prime }=0$ and find

\begin{equation}  \label{szx'}
\left<\left<\hat S_z\right>\right>_q = \frac{q x^{\prime} N}{4},
\end{equation}
that leads to an effective particle number

\begin{equation}
N_{eff}^0=qN,  \label{N0}
\end{equation}
as in \cite{nosotros}. Following the same mechanism and using (\ref{szx'}),
one finds that
\begin{equation}
Tr(\rho ^q)=2^{N(1-q)}.
\end{equation}

Remembering that $x^{\prime }=x/Tr(\rho ^q)$, it is possible to recover the
TMP normalized solution \cite{nosotros}

\begin{equation}
\left\langle \left\langle \hat{S}_z\right\rangle \right\rangle _q=\frac{qxN}4%
2^{N(q-1)},  \label{sz3}
\end{equation}
and
\begin{equation}
N_{eff}^{0\ ^{(3)}}=qN\ 2^{N(q-1)}.  \label{N03}
\end{equation}

For $x^{\prime }\rightarrow \infty ,$ and for $0<q<1,$ expression (\ref
{magne}) leads to an equation identical to that of \cite{nosotros}
\begin{equation}
\sum_{S=\delta }^{N/2}Y(S)\sum_{M=-S}^S\left( M-\left\langle \left\langle
\hat{S}_z\right\rangle \right\rangle _q\right) ^{1/(1-q)}=0,
\end{equation}
whose solution reads $\left\langle \left\langle \hat{S}_z\right\rangle
\right\rangle _q=N/2$.

\section{Conclusions}

In order to obtain the probability distribution $p_i$ that maximizes
Tsallis' entropy subject to appropriate constraints, Tsallis-Mendes-Plastino
extremize \cite{mendes}
\[
F=S_q-\alpha _0\left( \sum_{i=1}^wp_i-1\right) -\sum_{j=1}^n\lambda _j\left(
\frac{\sum_{i=1}^wp_i^qO_j^{(i)}}{\sum_{i=1}^wp_i^q}-\left\langle
\left\langle O_j\right\rangle \right\rangle _q\right) ,
\]
and obtain

\[
p_i=\frac{f_i^{1/(1-q)}}{\bar{Z}_q},
\]
where

\[
f_i=1-\frac{(1-q)\sum_j\lambda _j\left( O_j^{(i)}-\left\langle \left\langle
O_j\right\rangle \right\rangle _q\right) }{k\sum_jp_j^q},
\]
and $\bar{Z}_q$ is the partition function. Two rather unpleasant facts are
thus faced, namely,

\begin{itemize}
\item  $p_{i}$ explicitly depends upon the probability distribution
(self-reference).

\item  The Hessian of $F$ is not diagonal.
\end{itemize}

In this work we have devised a transformation from the original set of
Lagrange multipliers $\{\lambda _{j}\}$ to a new set $\{\lambda _{j}^{\prime
}\}$ such that

\begin{itemize}
\item  Self-reference is avoided.

\item  The Hessian of $F$ becomes diagonal.
\end{itemize}

As a consequence, all calculations, whether analytical or numerical, become
much simpler than in \cite{mendes}, as illustrated with reference to several
simple examples. The primed multipliers
\[
\lambda _j^{\prime }=\frac{\lambda _j}{\sum_i\,p_i^q}
\]
incorporate the $p_i$\footnote{%
that maximize the entropy} in their definition. Since one solves directly
{\it for the primed multipliers}, such a simple step considerably simplifies
the TMP treatment. Finally, we remark on the fact that the two sets of
multipliers lead to thermodynamical relationships that involve identical
intensive quantities (\ref{rel}).

\acknowledgements
The financial support of the National Research Council (CONICET) of
Argentina is gratefully acknowledged. F. Pennini acknowledges financial
support from UNLP, Argentina.


\begin{references}
\bibitem{mendes}  C. Tsallis, R. S. Mendes, and A. R. Plastino, {\it Physica
A} {\bf 261} (1998) 534.

\bibitem{t01}  C. Tsallis, {\it Braz. J. of Phys.} {\bf 29} (1999) 1, and
references therein. See also
http://www.sbf.if.usp.br/WWW\_pages/Journals/BJP/Vol129/Num1/index.htm

\bibitem{t1}  C. Tsallis, {\it Chaos, Solitons, and Fractals} {\bf 6} (1995)
539, and references therein; an updated bibliography can be found in
http://tsallis.cat.cbpf.br/biblio.htm

\bibitem{t03}  C. Tsallis, {\it Physics World 10} (July 1997) 42.

\bibitem{t04}  A. R. Plastino and A. Plastino, in {\it Condensed Matter
Theories}, Volume {\bf 11}, E. Lude\~{n}a (Ed.), Nova Science Publishers,
New York, USA, p. 341 (1996).

\bibitem{t05}  A. R. Plastino and A. Plastino, {\it Braz. J. of Phys.} {\bf %
29} (1999) 79.

\bibitem{katz}  E. T. Jaynes in {\it Statistical Physics}, ed. W. K. Ford
(Benjamin, NY, 1963); A. Katz, {\it Statistical Mechanics}, (Freeman, San
Francisco, 1967).

\bibitem{t2}  C. Tsallis, {\it J. Stat. Phys.} {\bf 52} (1988) 479.

\bibitem{aleman}  E. Fick and G. Sauerman, {\it The quantum statistics of
dynamic processes} (Springer-Verlag, Berlin, 1990).

\bibitem{pennini1}  F. Pennini, A. R. Plastino and A. Plastino, {\it Phys.
Lett. A} {\bf 208} (1995) 309.

\bibitem{pennini}  F. Pennini, A. R. Plastino and A. Plastino, {\it Physica A%
} {\bf 258} (1998) 446.

\bibitem{disisto}  R. P. Di Sisto, S. Mart\'{\i }nez, R. B. Orellana, A. R.
Plastino, A. Plastino, {\it Physica A} {\bf 265} (1999) 590.

\bibitem{renyi} A. R\`enyi, {\it Probability theory}
(North-Holland, Amsterdam, 1970).

\bibitem{beck}  C. Beck and F. Schl\"ogl, {\it Thermodynamics of chaotic systems}
(Cambridge University Press, Cambridge, England, 1993).




\bibitem{lenzi} E.K. Lenzi, R. S. Mendes and L. R. da Silva, {\it Physica A%
} (2000), in press.

\bibitem{t06}  A. Plastino and A. R. Plastino, {\it Braz. J. of Phys.} {\bf %
29} (1999) 50.

\bibitem{portesi}  M. Portesi, A. Plastino and C. Tsallis, {\it Physical
Review E }{\bf 52 }(1995) R3317.

\bibitem{nosotros}  S. Mart\'{\i }nez, F. Pennini, and A. Plastino, {\it %
Physica A} (2000), in press.
\end{references}
\end{document}